\documentclass{article}  
\usepackage{amsmath,amsfonts,amssymb}
\usepackage{graphicx}
\usepackage{fancyhdr}
\usepackage{rotating}
\usepackage{color}
\newcommand{\be}{\begin{equation}}
\newcommand{\ee}{\end{equation}}
\newcommand{\lo}{{\cal L}_0}

\title{Turbulence monitoring at Calern observatory with the Generalised Differential Image Motion Monitor} 
\author{Eric Aristidi$^a$, Yan Fant\'e\"{\i}-caujolle$^a$, Aziz Ziad$^a$, Julien Chab\'e$^b$, Christophe Giordano$^a$,\\ Catherine Renaud$^a$, Alohotsy Rafalimanana$^a$}
\date{\small\sl
$^a$Universit\'e C\^ote d'Azur, Observatoire de la C\^ote d'Azur, CNRS, laboratoire Lagrange, France\\
$^b$Universit\'e C\^ote d'Azur, OCA, CNRS, IRD, G\'eoazur, 2130 route de l'Observatoire, 06460 Caussols, France}

\begin{document}

  
  \maketitle 

\section*{Abstract}
The Generalised Differential Image Motion Monitor (GDIMM) was proposed a few years ago as a new generation instrument for turbulence monitoring. It measures integrated parameters of the optical turbulence, i.e the seeing, isoplanatic angle, scintillation index, coherence time and wavefront coherence outer scale. GDIMM is based on a fully automatic small telescope (28cm diameter), equipped with a 3-holes mask at its entrance pupil. The instrument is installed at the Calern observatory (France) and performs continuous night-time monitoring of turbulence parameters. In this communication we present long-term and seasonnal statistics obtained at Calern, and combine GDIMM data to provide quantities such as the equivalent turbulence altitude and the effective wind speed.

\section{INTRODUCTION}
\label{par:intro}  
The Generalized Differential Image Monitor (GDIMM) is an instrument designed to monitor integrated parameters of the atmospheric turbulence above astronomical observatories. It provides the seeing $\epsilon_0$, the isoplanatic angle $\theta_0$, the scintillation index $s_0$, the coherence time $\tau_0$ and the spatial coherence outer scale $\lo$. The 3 parameters $\epsilon_0$, $\theta_0$ and $\tau_0$ are of fundamental importance for adative optics (AO) correction. The scintillation has a strong impact on photometric signals from astronomical sources as well as on optical telecommunications with satellites. The outer scale ${\cal L}_0$ has a significant effect for large diameter telescopes (8m and above) and impacts low Zernike mode such as tip-tilt\cite{Winker91}. The GDIMM is at the moment the only monitor to provide simultaneously all these parameters.

GDIMM was proposed in 2014\cite{Aristidi14} to replace the old-generation turbulence monitor GSM\cite{Ziad00}. It is a compact instrument very similar to a DIMM\cite{Sarazinroddier90}, with 3 sub-apertures of different diameters. GDIMM observes bright single stars up to magnitude $V\sim 2$, at zenith distances up to 30$^\circ$. After a period of developpement and tests in 2013--2015, the GDIMM is operational since the end of 2015, as a part of the Calern atmospheric Turbulence Station (C\^ote d'Azur Observatory -- Calern site, France, UAI code: 010, Latitude=$43^\circ 45' 13''$~N, Longitude=$06^\circ 55' 22''$~E). GDIMM provides continuous monitoring of 5 turbulence parameters ($\epsilon_0$,  $s_0$, $\theta_0$,  $\tau_0$ and  $\lo$)  above the Calern Observatory. Data are displayed in real time through a website ({\tt cats.oca.eu}), as a service available to all observers at Calern. The other objective is that Calern becomes an operational on-sky test platform for the validation of new concepts and components in order to overcome current limitations of high angular resolution existing systems. Several activities regarding adaptive optics are operated at the M\'eO\cite{Samain08} and C2PU\cite{Bendjoya12} telescopes and they benefit of the data given by the CATS station.

The present communication is an update of previous papers\cite{Aristidi14, Aristidi18, Aristidi19}. We describe last
improvements and the present status of the instrument in Sect.~\ref{par:obs}. Long-term statistics (5 years) of turbulence above the Plateau de Calern are presented in Sect.~\ref{par:stats}.

\section{OBSERVATIONS AND DATA PROCESSING}
\label{par:obs}
GDIMM is based on a small commercial telescope (diameter 28cm) equipped with a 3 apertures pupil mask. 2 sub-pupils have a diameter of 6cm and are equipped with glass prisms oriented to give opposite tilts to the incident light. The 3rd sub-aperture is circular, its diameter is 10cm and it has a central obstruction of 4cm: this particular geometry is required to estimate the isoplanatic angle from the scintillation index\cite{Looshogge79}. At the focus of the telescope images are recorded by a fast camera allowing a framerate of 100 frames per second. Turbulence parameters are estimated on data cubes containing 2048 images. The first half of these cubes are taken with an exposure time of 5ms, the second half with 10ms, to allow compensation of the bias due to the exposure time. A set of parameter is calculated every 2mn and sent to a database accessible worldwide via a web interface ({\tt cats.oca.eu}). More details are given in previous papers\cite{Aristidi19}.

\begin{figure*}
\begin{center}
\includegraphics[width=70mm]{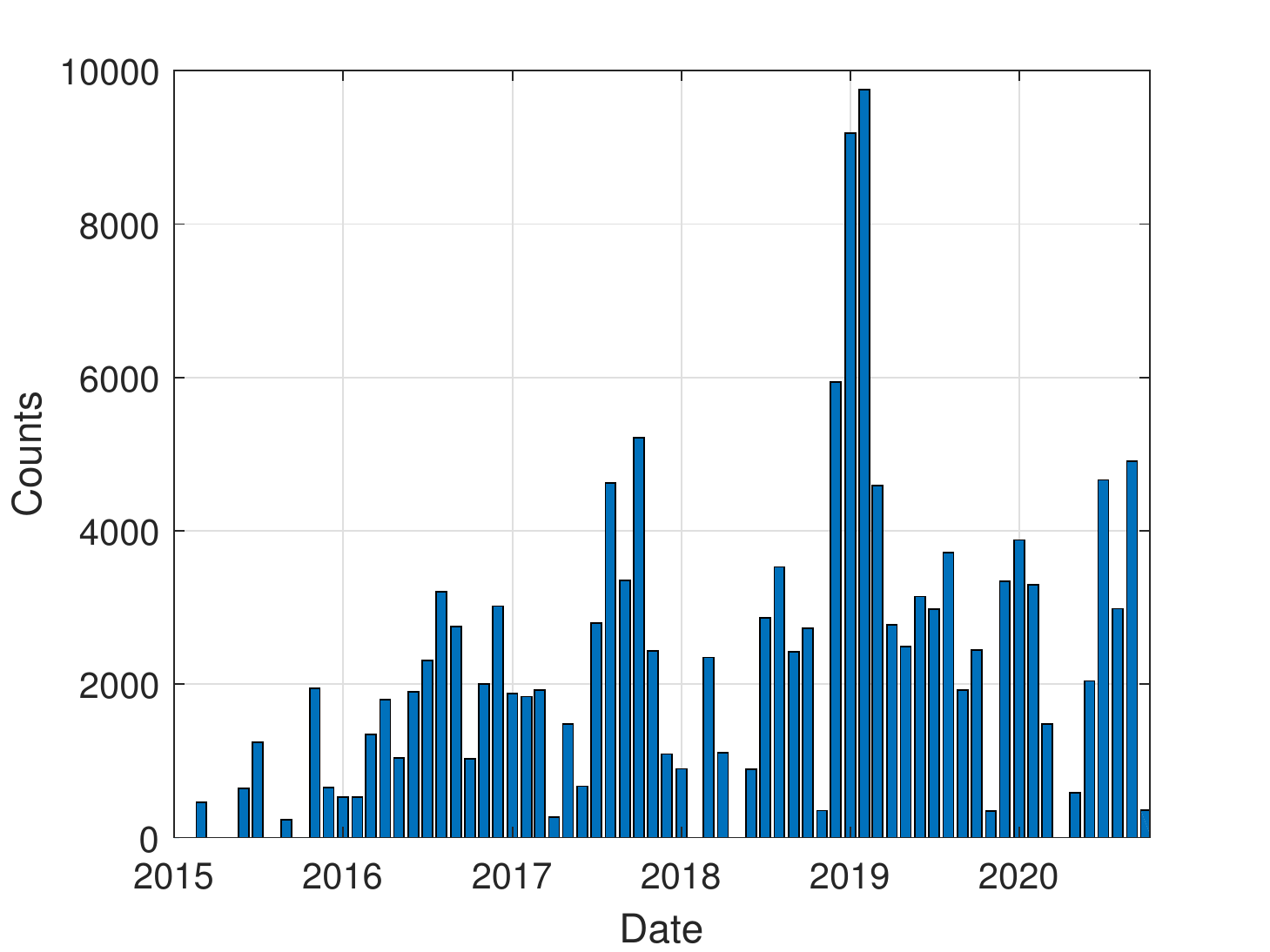}
\includegraphics[width=70mm]{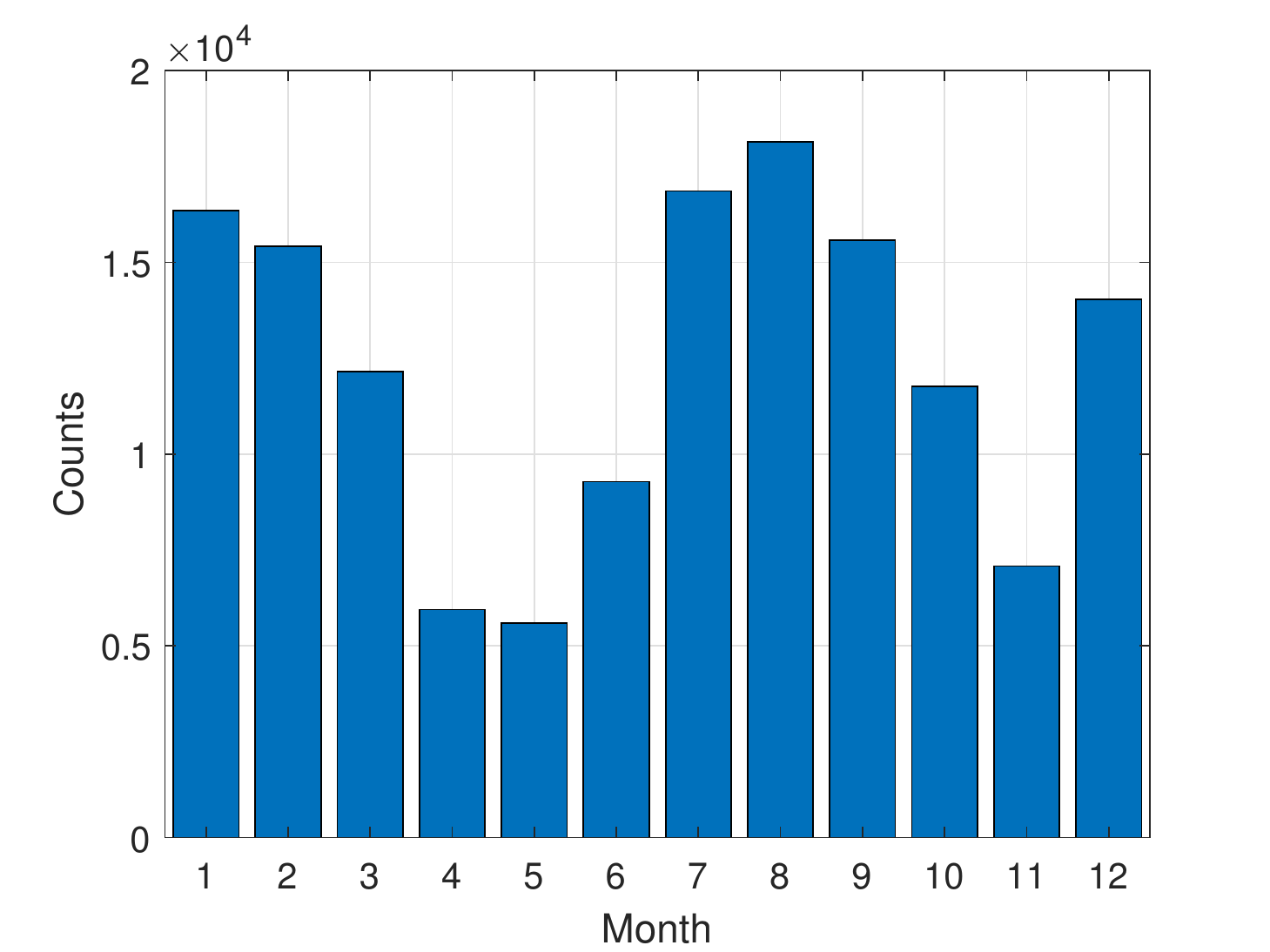}
\end{center}
\caption{Left: number of data points collected each month since 2015. Right: monthly number of data.}
\label{fig:nbdata}
\end{figure*}

The dataset analyzed here covers the period March 2015 to September 2020. The number of data points collected each month of the period is displayed in Fig.~\ref{fig:nbdata} (left). As the robotization of the instrument progressed during the years 2015--2016, the number of data increased. At the beginning of 2019, for a few months, we reduced the delay between successive acquisitions from 2 mn to 1mn, this resulted in a larger number of data (up to $\sim 10^4$ in Februrary 2019). Fig.~\ref{fig:nbdata} (right) shows the total number of measurements per month. Two periods in the year present a smaller number of data: April-May and November. They correspond to technical failures or adverse weather (in particular the spring 2018 was very bad and rainy, and the gap in April 2020 was due to the covid-19 lockdown).

\section{STATISTICS AT THE CALERN OBSERVATORY}
\label{par:stats}

\begin{table}
\begin{center}
\begin{tabular}{l|c|c|c|c|c|c|c}\hline
          & $\epsilon_0$  & $s_0$ & $\theta_0$  & $\tau_0$  &  $\lo$  & $\bar h$   & $\bar v$   \\
					&  [$''$] & [\%] & [$''$] &  [ms] &   [m] &  [m] &   [m/s] \\ \hline					
Median    				& 1.19  & 2.99 & 1.61  & 2.02  & 24.50 & 3394 & 13.31\\
Mean      				& 1.33  & 3.63 & 1.73  & 2.78  & 35.67 & 3645 & 13.84 \\
Std. dev.					 & 0.58 & 1.93 & 0.59  & 1.66  & 28.50 & 1473 & 5.37\\
$1^{st}$ quartile & 0.86  & 1.99 & 1.26  & 1.25  & 12.40 & 2520 & 9.71 \\
$3^{rd}$ quartile & 1.64  & 4.52 & 2.06  & 3.39  & 48.80 & 4511 & 17.08\\
$1^{st}$ centile & 0.46  & 0.60  & 0.66  & 0.44  & 2.70 & 1176 & 3.21\\
Last     centile & 3.55  & 11.78 & 3.94  & 13.13 & 141.50 & 8839 & 27.80\\ \hline
Paranal   			& 0.81 & 1.51 & 2.45 & 2.24 & 22 & 3256 & 17.3 \\
La Silla  			 & 1.64 & 4.63 &  1.25 & 1.46  & 25.5 & 3152  & 13.1  \\
Mauna Kea   		& 0.75 & 1.11 & 2.94 & 2.43 & 24 & 2931 &  17.2 \\ \hline
\end{tabular}
\end{center}
\caption{Statistics of turbulence parameters measured at Calern (at the wavelength $\lambda=0.5\mu$m) during the period March 2015--Sept 2020. Paranal, La Silla and Mauna Kea values are from the GSM database.}
   \label{table:paramstat}
	\end{table}

\begin{figure}
\begin{center}
\includegraphics[width=17cm]{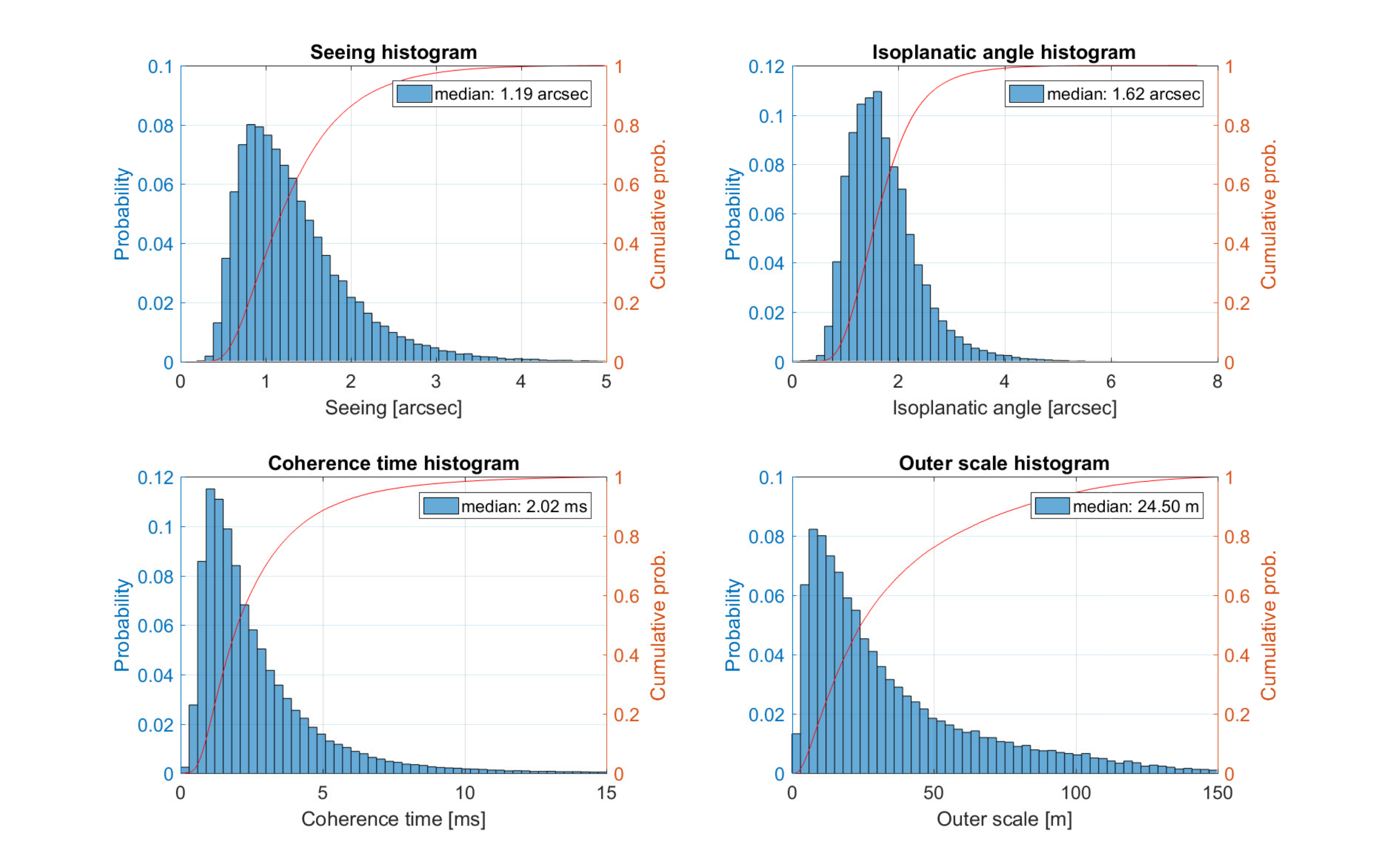}
\caption{Histograms of the seeing, isoplanatic angle, coherence time and outer scale at Calern, calculated at the wavelength $\lambda=0.5\mu$m for the period June 2015--September 2020.}
   \label{fig:paramshisto}
\end{center}
\end{figure}

We collected about 148\,000 turbulence parameter sets during the 5~year period. The number for the outer scale ${\cal L}_0$ is lower (54\,000): this parameter is sensitive to telescope vibrations and a reliable value is not always available\cite{Aristidi19}. Statistics are presented in Table~\ref{table:paramstat} for the 5 turbulence parameters ($\epsilon_0$, $s_0$, $\theta_0$ $\tau_0$, $\lo$). Note that the scintillation index $s_0$ is measured through a 10cm diameter sub-pupil: this causes a low-pass spatial filtering and values are lower than the actual scintillation defined for a zero-diameter pupil\cite{Roddier81}.

We also provide values for the equivalent turbulence altitude $\bar h$ and the effective wind speed $\bar v$ (weighted average of the wind speed over the whole atmosphere). A definition of these quantities can be found in papers by Roddier\cite{Roddier81, Roddier82}, and we estimated them by the following equations:
\be
\bar{v}=0.31 \frac{r_0}{\tau_0}
\ee
and
\be
\bar{h}=0.31 \frac{r_0}{\theta_0}
\ee

Histograms are displayed in Fig.~\ref{fig:paramshisto} and show a classical log-normal shape for all parameters. We did not display the scintillation histogram since is is directly deduced from $\theta_0$ by an exact analytic relation. A comparison with other astronomical sites in the world (examples for Paranal, La Silla and Mauna Kea are given in Table~\ref{table:paramstat}) show that the Calern plateau is an average site.

\begin{figure}
\begin{center}
\includegraphics[width=17cm]{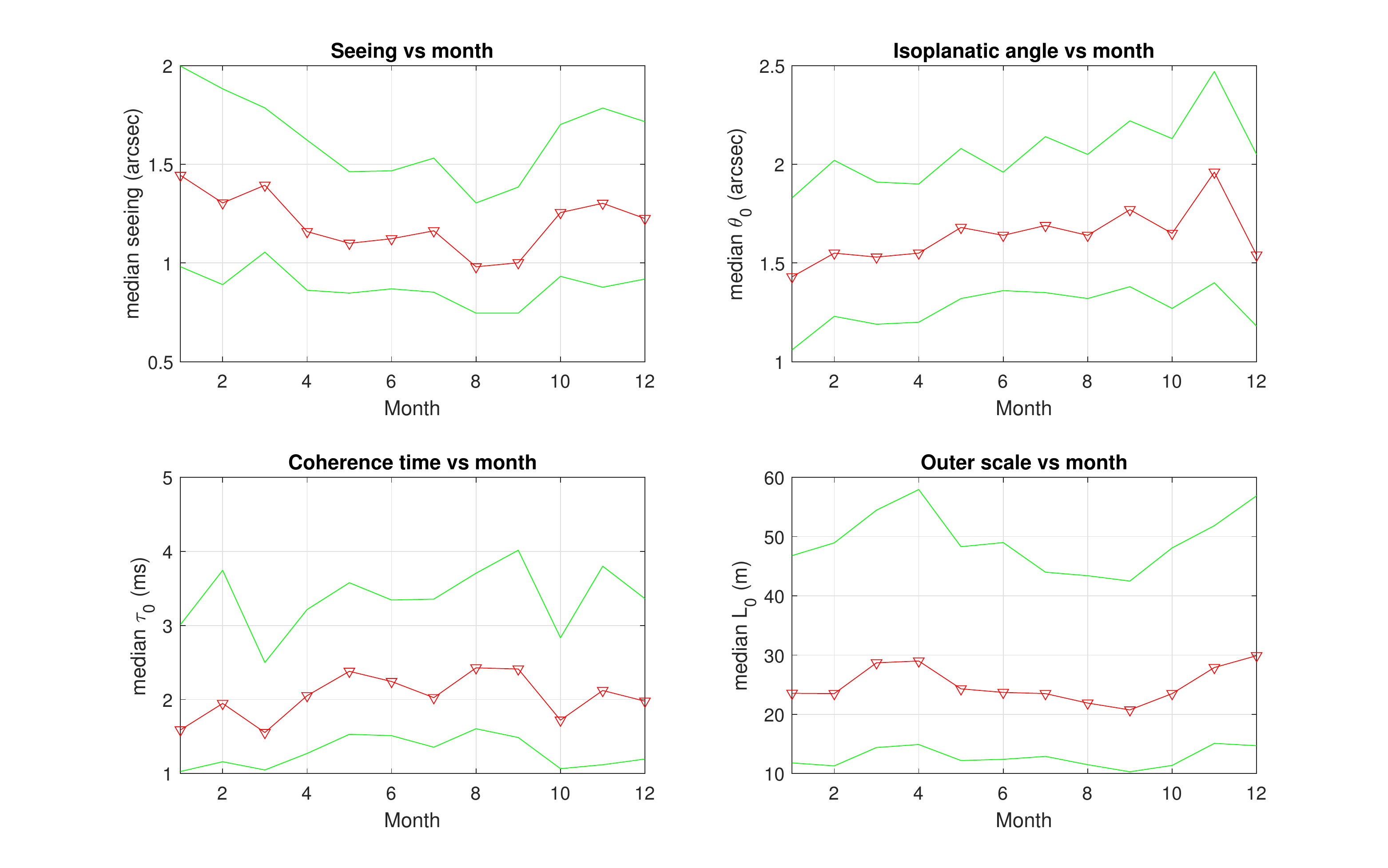}
\caption{Monthly median values of the four parameters ($\epsilon_0$, $\theta_0$, $\tau0$, $\lo$), calculated at the wavelength $\lambda=0.5\mu$m for the period June 2015--September 2020. On each graph, the red curve is the median value, green curves are the first and third quartiles.}
   \label{fig:params_vs_month}
\end{center}
\end{figure}

\begin{figure}
\begin{center}
\includegraphics[width=8cm]{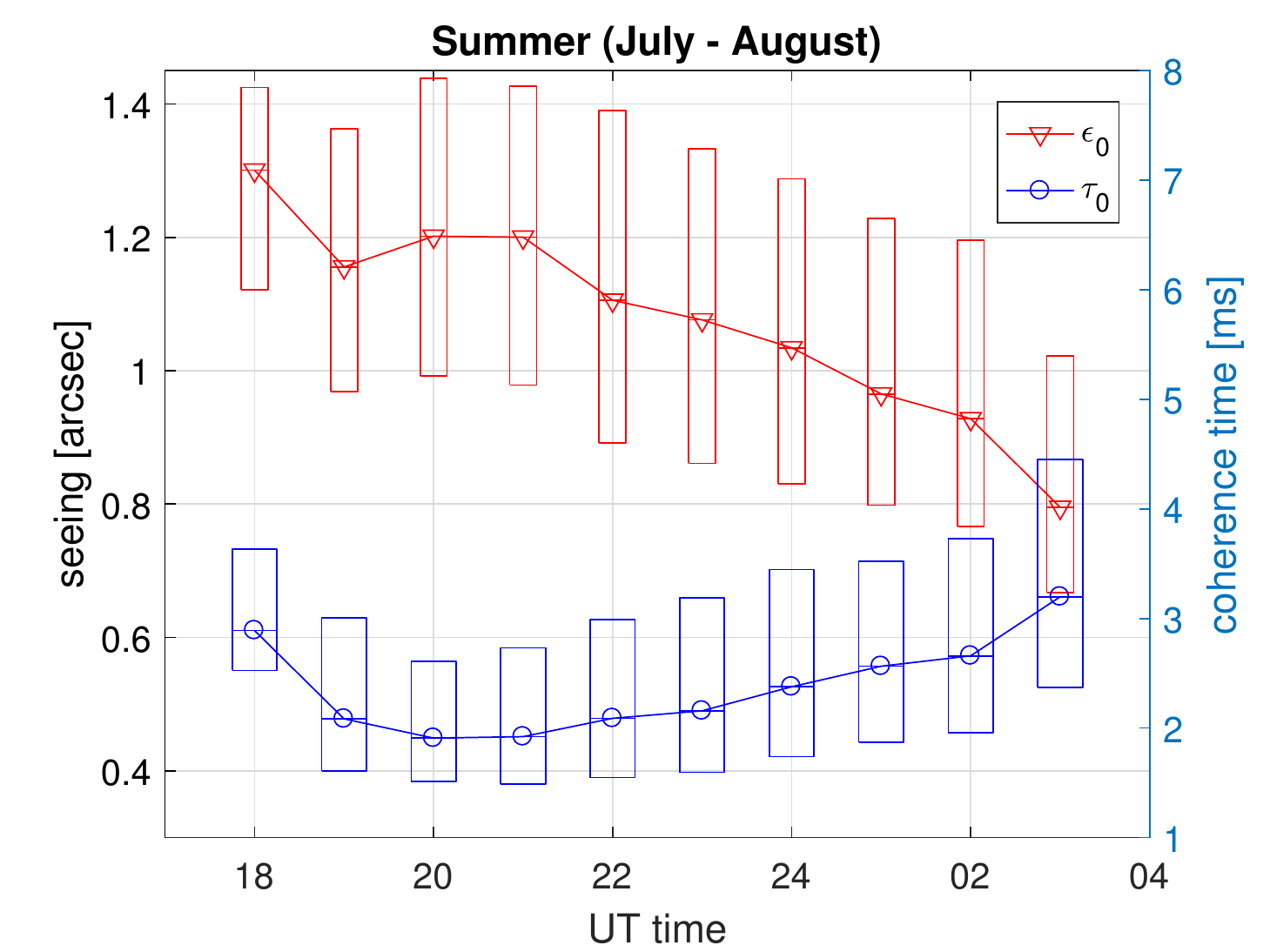}
\caption{Hourly variations of the seeing $\epsilon_0$ and coherence time $\tau_0$ during the summer (July-August). Symbols (triangle/circle) are the median values. Error rectangles lower and upper edges are the 1st and 3rd quartiles. The data set represents a total of 27300 measurements for 5 summers (2015 to 2019).}
   \label{fig:hourlyseeingtau0}
\end{center}
\end{figure}

Fig.~\ref{fig:params_vs_month} displays the monthly evolution of parameters. The seeing is slightly lower in summer, we measured a median value of $1.06''$ in July and August (the median winter seeing during the period November--March is 1.34$''$). As a consequence, the median coherence time is higher in summer (2.24ms in July--August, 1.78ms in November--March). Also, Fig.~\ref{fig:hourlyseeingtau0} shows that, in Summer, there is a dependence of  $\epsilon_0$ and $\tau_0$ with time. The median seeing decreases below 0.8$''$ at the end of the night (while the coherence time increases). Nothing similar was observed during the winter. The isoplanatic angle and the outer scale didn't show any noticeable time dependence, both in summer and  winter. The outer scale $\lo$ has values similar to other sites such as Mauna Kea or La Silla.

\begin{figure}
\begin{center}
\includegraphics[width=15cm]{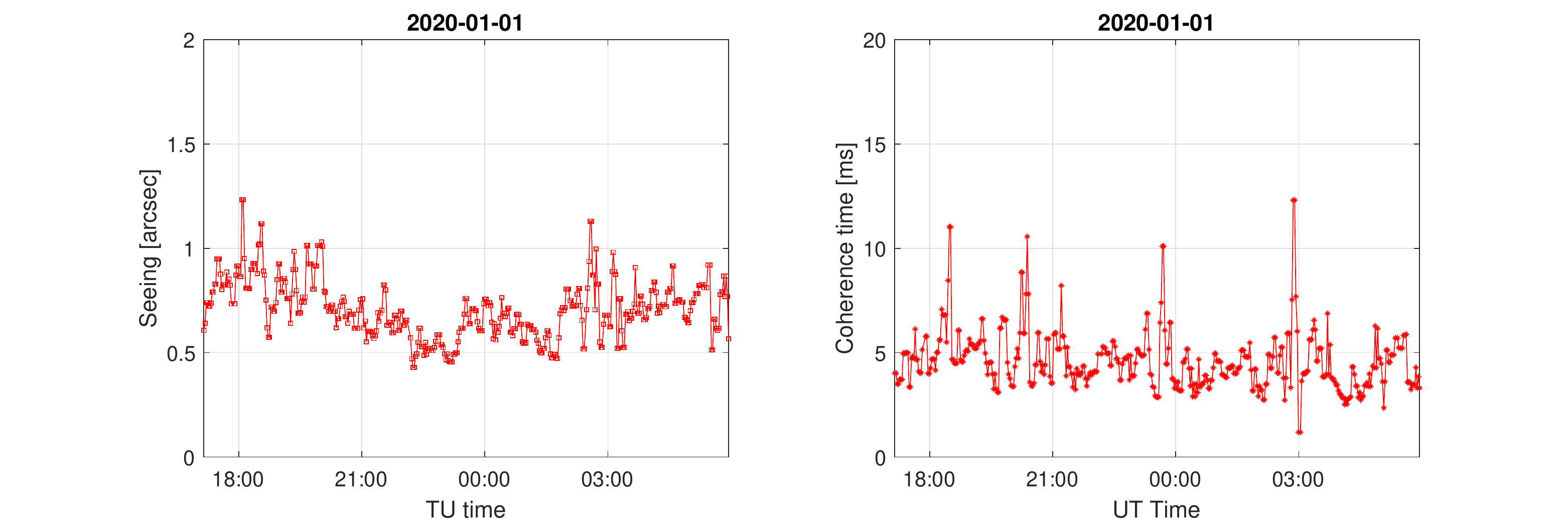}
\caption{Exceptional conditions observed on Jan. 1st, 2020. Left: seeing time series (the median value for this night was 0.70$''$), right: coherence time (median value 4.3ms).}
   \label{fig:seeingtsgood}
\end{center}
\end{figure}

Sequences of several hours of good seeing were sometimes observed, which is a good point for this site (and already known by ``old'' observers on interferometers during the 80's and 90's). An exemple is shown on Fig.~\ref{fig:seeingtsgood}. It is a time series recorded on Jan. 1st, 2020, showing a seeing below 1$''$ during 12~hours, the median seeing for that night was 0.69$''$.

\begin{figure*}
\begin{center}
\includegraphics[width=8cm]{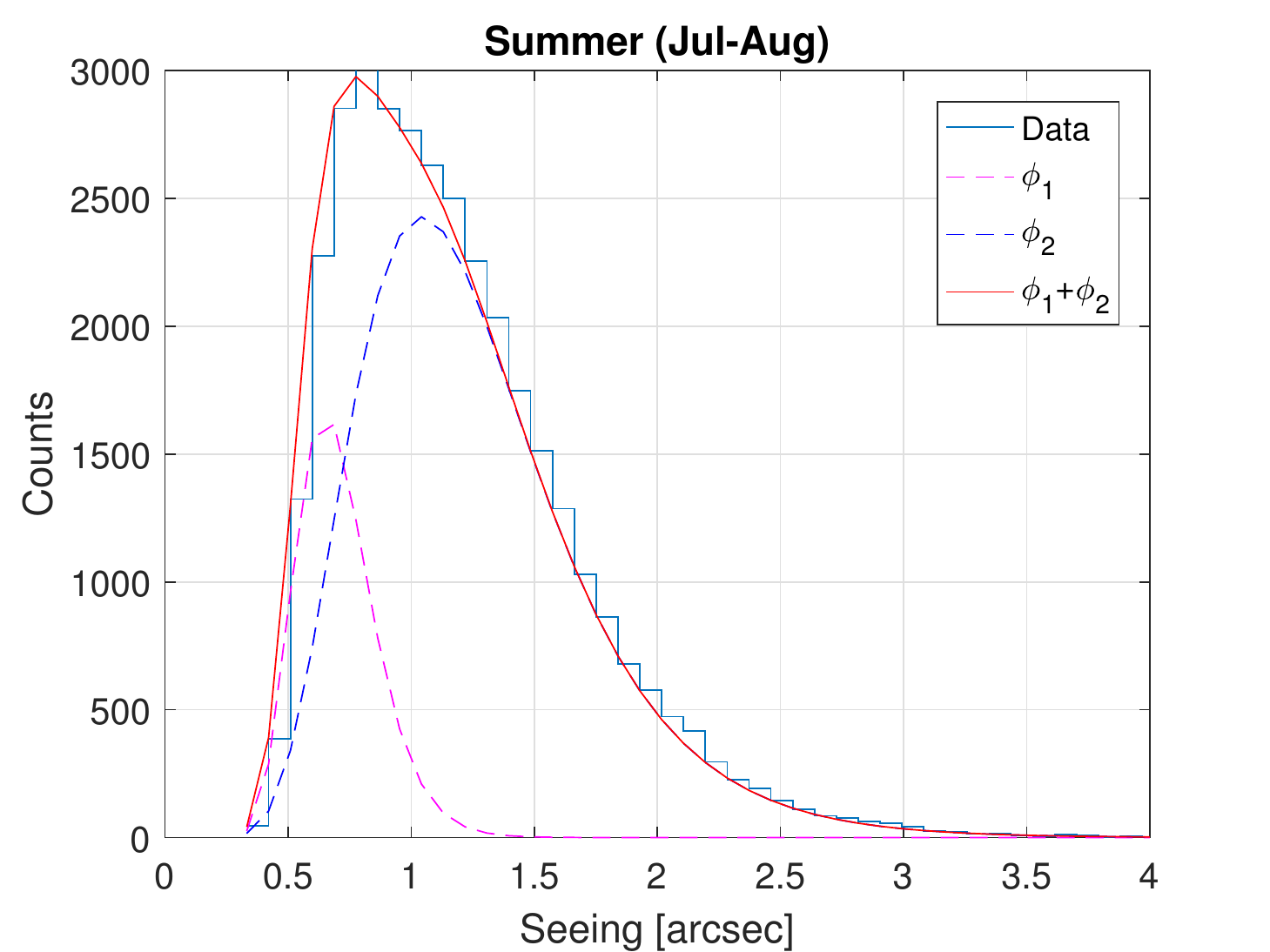}
\includegraphics[width=8cm]{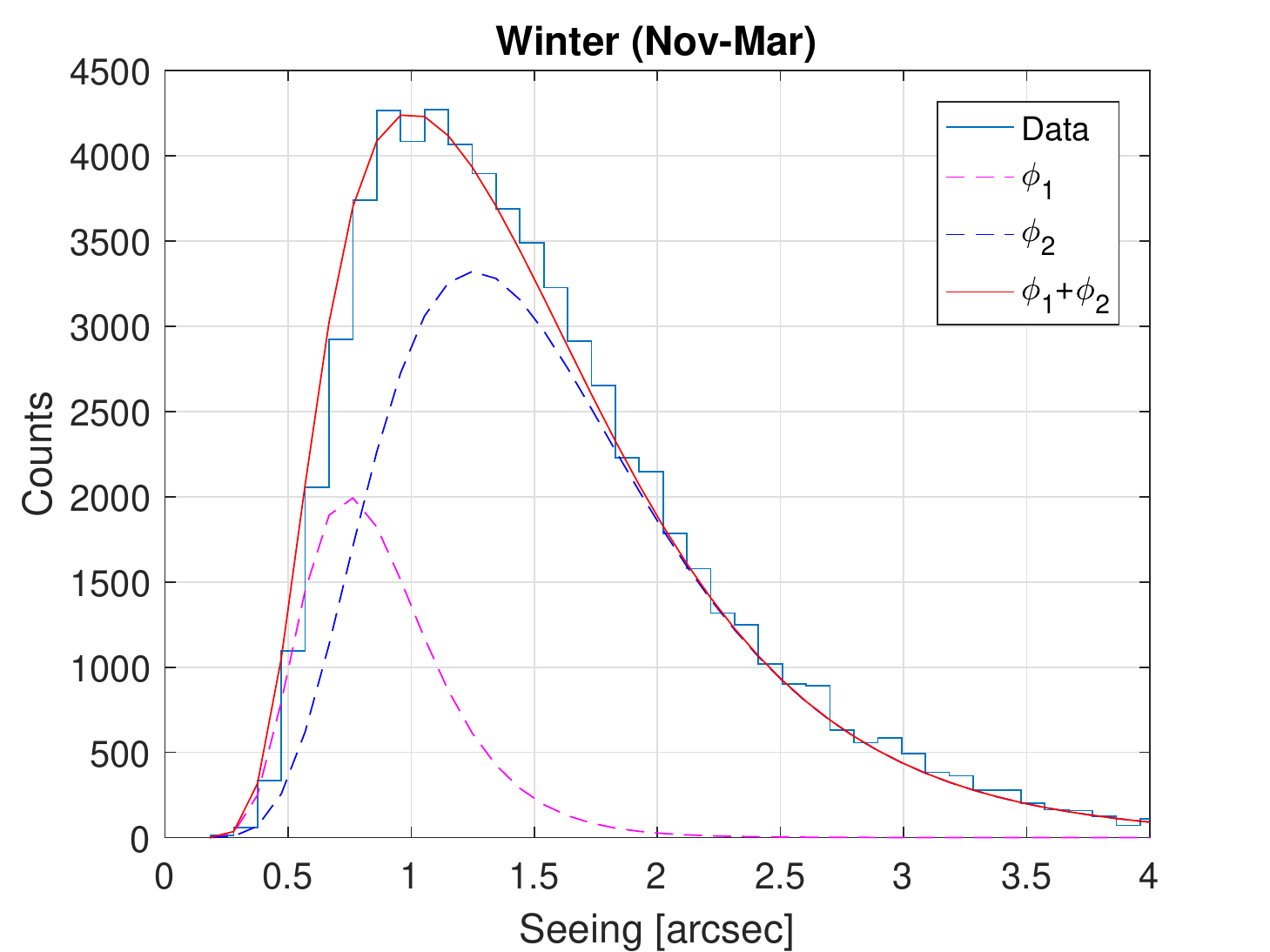}
\caption{Left: seeing histogram for the summer (July--August). Right: seeing histogram for the winter (Nov--March). Superimposed curves $\phi_1$ and $\phi_2$ are a least-square fit by a sum of two log-normal distributions (individual log-normal curves are dashed lines).}
   \label{fig:histseeingfit}
\end{center}
\end{figure*}

Fig.~\ref{fig:histseeingfit} displays seasonal seeing histograms, calculated for the summer (July and August) and the winter (November--March). They appear to be well modelled by a sum of two log-normal functions (they appear as dashed curves on the plots, their sum is the solid line). This is an evidence of the existence of two regimes: a ``good seeing'' distribution $\phi_1$ with a median value $\epsilon_1$ and a ``medium seeing'' distribution $\phi_2$ with a median value $\epsilon_2$. In summer, we have $\epsilon_1=0.68''$ (the good seeing distribution contains 20\% of the data) and $\epsilon_2=1.04''$ (80\% of the data). In winter we have $\epsilon_1=0.76''$ (21\% of the data) and  $\epsilon_2=1.25''$ (79\% of the data). The time series displayed in Fig.~\ref{fig:seeingtsgood} corresponds to a realisation of the good seeing distribution, during 12~hours.

\subsection*{Temporal stability of parameters}
The stability of turbulence parameters is an important concern. Temporal fluctuations were studied by different authors\cite{Racine96, Ziad99}. The characteristic time of stability is generally of a few minutes for temperate sites (e.g. Mauna Kea and La Silla).

We define here an interval of stability as a continuous period of time $t_s$ in which a parameter is {\em better} than a given threshold $x_0$. The term {\em better} means {\em lower} for the seeing, as low values of the seeing correspond to high Strehl ratio. It means {\em higher} for the coherence time and isoplanatic angle. In this interval, we allow the parameter to be better than $x_0$ during a few minutes (we took 4 mn, i.e. two sampling intervals of parameter time series). For a given value of $x_0$, we calculate the distribution of $t_s$ on the whole dataset. Its median value gives the characteristic time of stability. This approach was the one used for the site characterization of Dome C by means of DIMMs\cite{Aristidi09, Fossat10} and Single Star Scidar\cite{Giordano12}. 

\begin{figure*}
\begin{center}
\includegraphics[width=60mm]{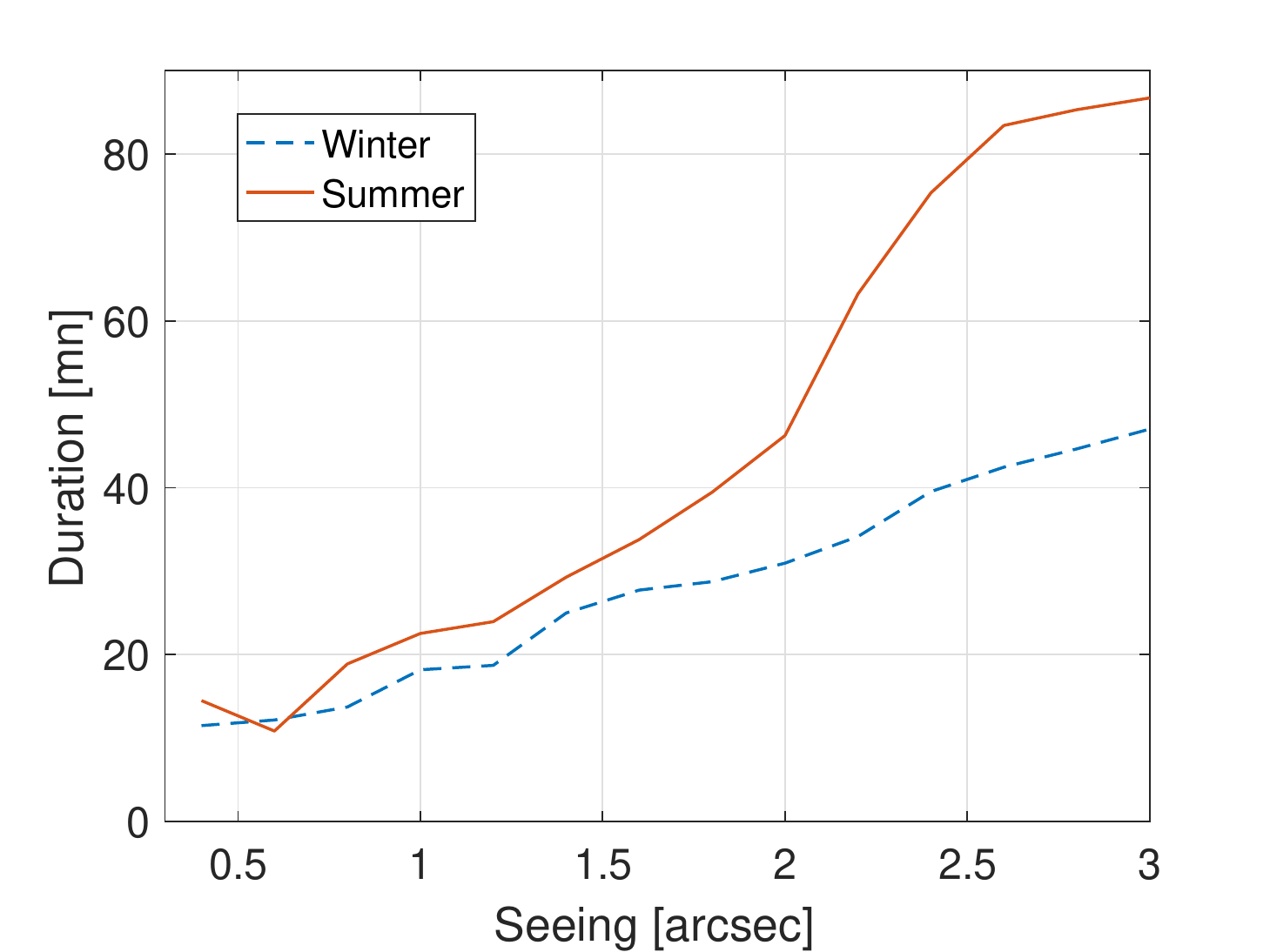}\hskip -5mm
\includegraphics[width=60mm]{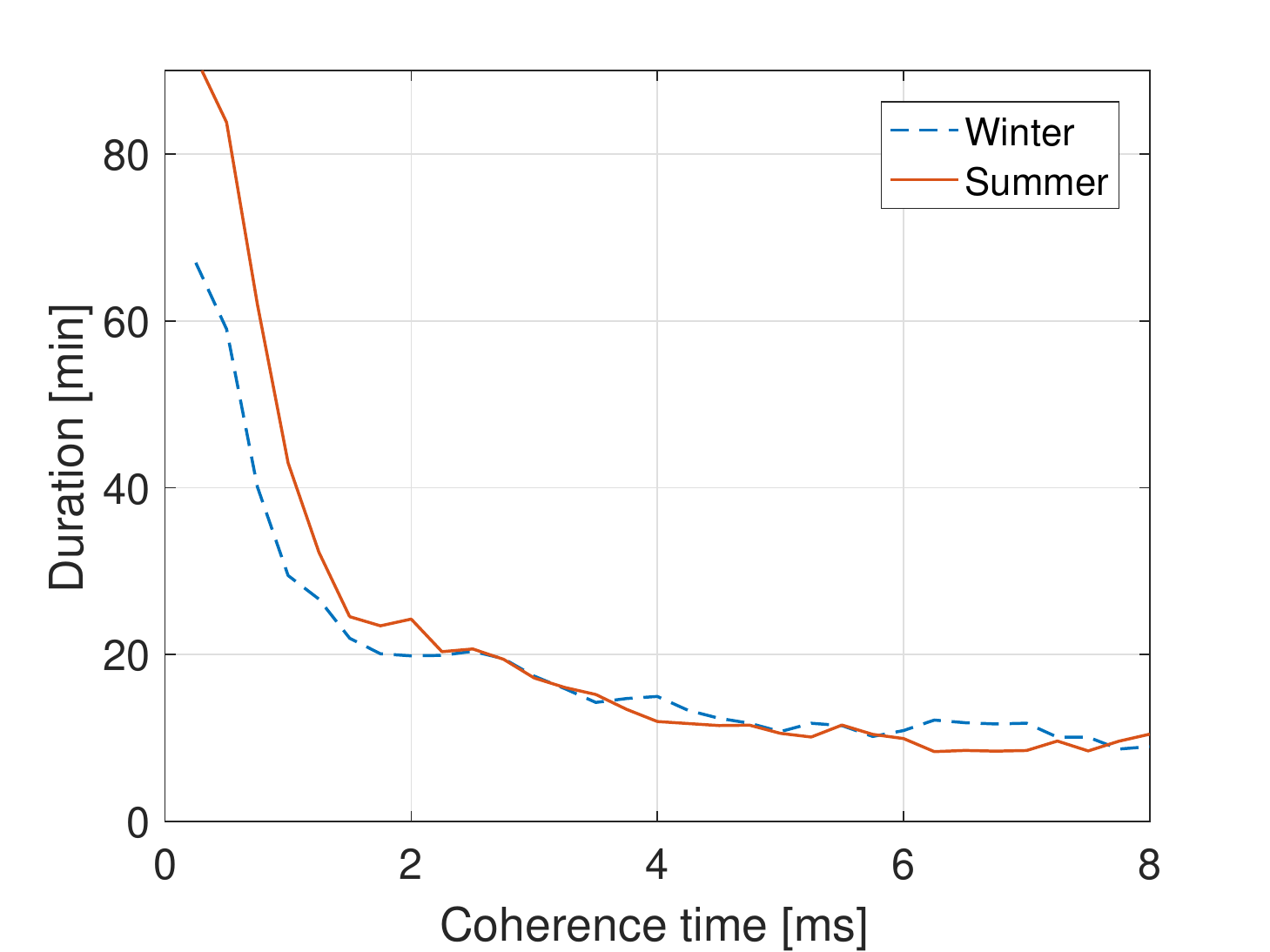}\hskip -5mm
\includegraphics[width=60mm]{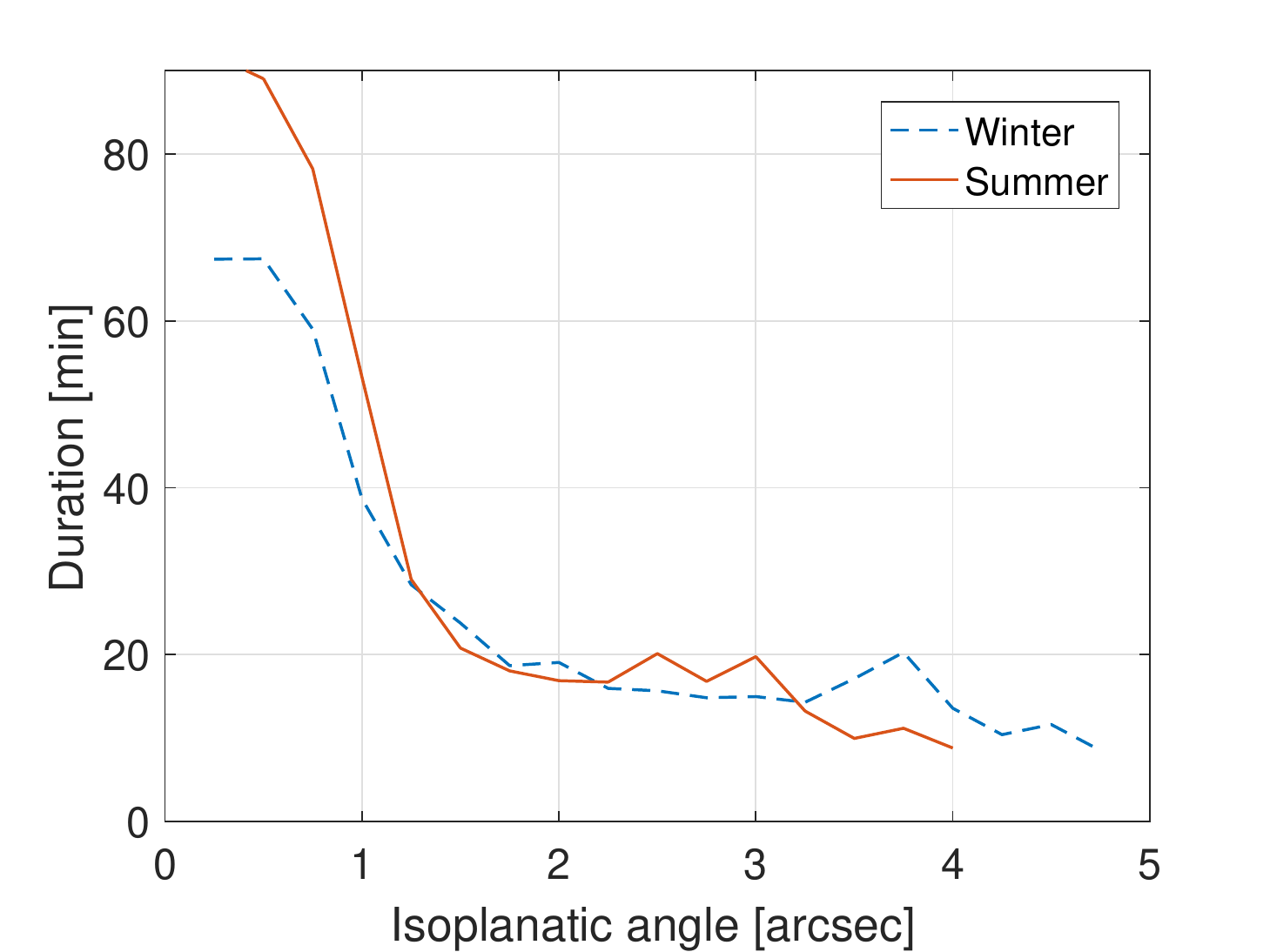}
\caption{Temporal stability of parameters, given by the continuous time a parameter is better than a given value (``better'' means lower for the seeing, and higher for the coherence time and the isoplanatic angle). Curves are plotted for the summer (July--August) and winter (November to Mars) datasets.}
   \label{fig:stability}
\end{center}
\end{figure*}

We performed this analysis for the 3 parameters $\epsilon_0$, $\tau_0$ and $\theta_0$, for the summer and the winter periods. We could not obtain results for the outer scale, since it is not always measurable because of telescope vibrations (data sequences show large gaps that hamper the calculation of $t_s$). Results are displayed in Fig~\ref{fig:stability}. For the seeing, the characteristic time increases with the seeing value and saturates. It is of the order of 20mn for seeing values around $1''$, and is longer during the summer. For $\tau_0$ and $\theta_0$ we observe that $t_s$ decrease with the threshold $x_0$, down to a value of $\sim$10mn. Here again, summer conditions appear to be slightly more stable.

Note that the curves are likely to be biased by interruptions of the observations (change of target star, clouds, day-night cycle). But the mean value of uninterrupted sequences duration is $\sim 150$mn, much larger than the characteristic stability times.

\section{CONCLUSION}
We have presented statistics of optical turbulence parameters above the plateau de Calern, measured for five years with the GDIMM monitor. GDIMM is a part of the CATS station which aims at monitoring atmospheric turbulence parameters and vertical profiles on the site of Calern. The station is fully automatic, using informations from a meteo station and an All-Sky camera to replace human interventions.

Seeing and isoplanatic measurements are given by differential motions and scintillation, according to well known techniques. The coherence time is derived from AA structure functions. The method is simple and has proven to give satisfactory results\cite{Ziad12}. It may be used as well with a classical DIMM providing that the acquisition camera is fast enough (typically 100 frames/sec) to properly sample the AA correlation times. The outer scale is derived from ratios of absolute to relative AA motions\cite{Aristidi19}. This parameter is the most sensitive to telescope vibrations, and requires a good stability of the telescope mount and pillar. 

Simultaneous observations of GDIMM and another instrument, the Profiler of Moon Limb, which estimates the vertical profile of turbulence, gave concordant results\cite{Ziad19, Chabe20}. A portable version of the GDIMM has been developped
in parallel to the Calern one, to perform turbulence measurements at any site on the world. Discussions with the ESO
(European Southern Observatory) are currently in progress to make observations at Paranal and Armazones.

\section*{Acknowledgments}
%
%
This CATS project has been done under thefinancial support of CNES, Observatoire de la C\^ote d'Azur,
Labex First TF, AS-GRAM, Federation Doeblin, Universit\'e de Nice-Sophia Antipolis and Re\'gion Provence Alpes
C\^ote d'Azur. We would like thanks all the technical and administrative staff of the Observatoire de la C\^ote
d'Azur and our colleagues from the Astrog\'eo team from G\'eoAzur laboratory for their help and support all along
the project namely Pierre Exertier, Etienne Samain, Dominique Albanese, Mourad Aimar, Jean-Marie Torre,
Emmanuel Tric, Thomas Lebourg and Sandrine Bertetic.

\bibliography{biblio}   
\bibliographystyle{spiebib}   

\end{document}